\documentclass[12pt]{iopart}

\usepackage{iopams}  
\usepackage[english]{babel}
\usepackage{graphicx}
\usepackage{color}
\usepackage{bbm}
\usepackage{cite}

\begin{document}

\title[A classical simulation of nonlinear JC and Rabi models ]{A classical simulation of nonlinear Jaynes--Cummings and Rabi models in photonic lattices}

\author{B. M. Rodr\'{\i}guez-Lara$^{\ast 1}$, Francisco Soto-Eguibar$^{1}$, Alejandro Z\'arate C\'ardenas$^{1}$, and H. M. Moya-Cessa$^{1,2}$}

\address{$^{1}$Instituto Nacional de Astrof\'{\i}sica, \'Optica y Electr\'onica,\\ Calle Luis Enrique Erro No. 1, Sta. Ma. Tonantzintla, Pue. CP 72840, M\'exico.\\ $^{2}$CREOL/Collefe of Optics, University of Central Florida, Orlando, Florida, USA.}

\ead{$^{\ast}$bmlara@inaoep.mx}

\date{\today}
\begin{abstract}
The interaction of a two-level atom with a single-mode quantized field is one of the simplest models in quantum optics.
Under the rotating wave approximation, it is known as the Jaynes-Cummings model and without it as the Rabi model.
Real-world realizations of the Jaynes-Cummings model include cavity, ion trap and circuit quantum electrodynamics.
The Rabi model can be realized in circuit quantum electrodynamics. 
As soon as nonlinear couplings are introduced, feasible experimental realizations in quantum systems are drastically reduced. 
We propose a set of two photonic lattices that classically simulates the interaction of a single two-level system with a quantized field under field nonlinearities and nonlinear couplings as long as the quantum optics model conserves parity. 
We describe how to reconstruct the mean value of quantum optics measurements, such as photon number and atomic energy excitation, from the intensity and from the field, such as von Neumann entropy and fidelity, at the output of the photonic lattices.
We discuss how typical initial states involving coherent or displaced Fock fields can be engineered from recently discussed Glauber-Fock lattices. 
As an example, the Buck-Sukumar model, where the coupling depends on the intensity of the field, is classically simulated for separable and entangled initial states.
\end{abstract}


%
\section{Introduction}

Optical analogy has been of great importance to the development of quantum mechanics; cf. \cite{Kemble1958}. 
Quantum information processing has been one of the fields where authors have extensively studied classical analogies by use of linear optics elements involving classical light beams \cite{Spreeuw1998p361} or single-photons \cite{Cerf1998p1477}.
In recent years, the advent of technology that allows manufacturing arrays of optical waveguides has detonated the classical simulation of quantum and relativistic systems \cite{Longhi2006p110402,Perets2008p170506,DellaValle2008p011106,Bromberg2009p253904,Dreisow2009p076802,Longhi2009p243,Lahini2010p163905,Dreisow2010p143902,Longhi2010p075102,Longhi2011p3248,Longhi2011p453,Longhi2012p435601,Longhi2012p012112,Garanovich2012,PerezLeija2013p012309}.

We are interested in the classical simulation of a single two-level atom interacting with a quantized field, the basic building block in quantum optics, under nonlinear processes in the field and couplings as the linear interaction has already been classically simulated \cite{Longhi2011,Crespi2012p163601}.
The quantum model of interaction between an ensemble of two-level atoms and a radiation field under a series of approximations was introduced by Dicke \cite{Dicke1954p99}.
The exact solution for the single atom case was found by Jaynes and Cummings \cite{Jaynes1963p89} and the Jaynes-Cummings (JC) model became a work horse of quantum optics  as well as the Dicke model. 
Among other interesting phenomena, the JC model shows a collapse and revival of the mean atomic excitation energy when interacting with coherent radiation \cite{Eberly1980p1323}, but it has not been possible to describe it in closed form.
Buck and Sukumar (BS) presented a JC model with nonlinear coupling where the mean atomic excitation energy can be evaluated in exact closed form and exhibits a periodic collapse and revival similar, for small times, to that seen in the JC model \cite{Buck1981p132}.
The most general JC model involving nonlinear couplings, up to our knowledge, was presented and solved by Kochetov \cite{Kochetov1987p2433}, 
\begin{eqnarray}
\hat{H} = \omega_{f} \hat{a}^{\dagger} \hat{a} + \omega_{0} \hat{S}_{z} + \lambda \left[ \hat{a}^{\dagger k} \hat{a}^{l} f(\hat{a}^{\dagger} \hat{a}) \hat{S}_{+} + f(\hat{a}^{\dagger} \hat{a}) \hat{a}^{\dagger l} \hat{a}^{k}  \hat{S}_{-} \right].
\end{eqnarray}
The Kochetov model describes the interaction between a single atomic system, with a finite number of equidistant levels, and a single-mode radiation field. 
The atom is described by the generators $\hat{S}_{i}$ with $i=z,+,-$ that obey an $su(2)$ algebra, $\left[\hat{S}_{z},\hat{S}_{\pm} \right] = \pm \hat{S}_{\pm}$ and $\left[\hat{S}_{+},\hat{S}_{-} \right] = 2 \hat{S}_{z}$, and the inter-level energy is given by $\omega_{0}$. The field is described by the creation (annihilation) operators, $\hat{a}^{\dagger}$($\hat{a}$), and the frequency $\omega_{f}$. The parameter $\lambda$ is a coupling constant and the function $f(\hat{a}^{\dagger} \hat{a})$ is a real well-behaved function of the number operator.
Peculiar phenomena has been found among the years with specific realizations of the Kochetov model; e.g. for just two levels \cite{Schlicher1989p97, Phoenix1990p116, Benivegna1994p353, Vogel1995p4214, MatosFilho1996p4560, Yang1997p4545,RodriguezLara2013p87}.
Some other nonlinear models studied along the years for the two-level case involve: driving \cite{Joshi2000p043812}, field nonlinearities \cite{AbdelAty2002p37,Cordero2011p135502,delosSantosSanchez2012p015502}, or consider the coupling without the rotating wave approximation (RWA) \cite{Naderi2005p397}.
The great majority of these models does not have a feasible experimental realization with current quantum technology.

In this contribution, we are interested in a specific class of the Kochetov model that we extend to consider field nonlinearities and the counter rotating terms neglected by the RWA. 
We present our nonlinear Rabi model and its characteristics in the following section; in short, it conserves parity and this allows us to split the Hilbert space in two parity basis.
Then, we show that these two parity basis allow us to describe the dynamics of our quantum model with classical light propagating in a set of two photonic lattices.
We show how to construct typical initial states; e.g. an initial state involving just a photon number state and the ground or excited state corresponds to light impinging just one waveguide and those involving coherent states can be prepared through Glauber-Fock photonic lattices \cite{PerezLeija2010p2409, RodriguezLara2011p053845, PerezLeija2012p013848}.
We also discuss the most common quantum optics measurements and show how they are related to measurements of the intensity at the end of the classical simulators, with just an exception that requires knowledge of the amplitude and phase at each output. 
In order to provide a working example, we present the BS model in Section 4 and close with a brief conclusion.

\section{The quantum optics model}

We are interested in a general nonlinear Rabi model, 
\begin{eqnarray} \label{eq:OurH}
\hat{H} &=& h(\hat{n}) + \frac{\omega_{0}}{2} \hat{\sigma}_{z} + g_{-} \left( \hat{a} \frac{f(\hat{n})}{\sqrt{\hat{n}}} \hat{\sigma}_{+} +  \frac{f(\hat{n})}{\sqrt{\hat{n}}} \hat{a}^{\dagger} \hat{\sigma}_{-}   \right) + \nonumber \\ &&  + g_{+} \left( \hat{a} \frac{f(\hat{n})}{\sqrt{\hat{n}}} \hat{\sigma}_{-} +  \frac{f(\hat{n})}{\sqrt{\hat{n}}} \hat{a}^{\dagger} \hat{\sigma}_{+}   \right),
\end{eqnarray}
with well-behaved real functions $h(\hat{n})$ and $f(\hat{n})$ in terms of the number operator, $\hat{n} = \hat{a}^{\dagger} \hat{a}$, and where the operators $\hat{\sigma}_{i}$ with $i=z,+,-$ are Pauli matrices. 
We have split the nonlinear Rabi coupling $\left( \hat{a} ~f(\hat{n})/\sqrt{\hat{n}} + f(\hat{n})/\sqrt{\hat{n}} ~\hat{a}^{\dagger} \right) \left(\hat{\sigma}_{-}  + \hat{\sigma}_{+}  \right)$ into Jaynes-Cummings coupling described by the coupling parameter $g_{-}$ and counter-rotating terms described by the parameter $g_{+}$ for reasons that will become apparent in the next section. 
As mentioned before, this general Hamiltonian is not physical realizable with current experimental setups. 
For example, in order to implement (\ref{eq:OurH}) in cavity- or circuit-QED the function $h(\hat{n})$ must be linear to account for the free-field energy or quadratic at most to describe the effect of a Kerr medium, $h(\hat{n}) = \omega \hat{n} + \kappa \hat{n}^{2}$, and the coupling function $f(\hat{n})/ \sqrt{n}$ must be a constant.
Spin chains \cite{Crisp1993}, ion traps \cite{deMatosFilho1996p608,MoyaCessa2000p025401,MoyaCessa2012p229}, and atoms in optical lattices \cite{Esslinger2010p129}, just to mention a few examples, may provide a way to realize some of the Hamiltonians covered by this class as far as one could attain precise control of the size, self-energies and couplings in the spin chain analog, drive an ion with multiple lasers in the second, and engineer adequate optical lattices in the latter.

The Hamiltonian in (\ref{eq:OurH}) has constant of motion: parity, which can be  defined as
\begin{eqnarray}
\hat{\Pi} = - (-1)^{\hat{n}} \hat{\sigma}_{z};
\end{eqnarray}
that is $\left[\hat{H}, \hat{\Pi} \right]=0$. 
Conservation of parity allows us to define two orthogonal parity bases,
\begin{eqnarray}
\vert +, j \rangle &=& \left( \hat{B}^{\dagger} \right)^{j} \vert 0, g \rangle =\left\{ \vert 0,g \rangle, \vert 1,e \rangle, \vert 2,g \rangle, \ldots   \right\}, \\
\vert -, j \rangle &=& \left( \hat{B}^{\dagger} \right)^{j} \vert 0, e \rangle = \left\{ \vert 0,e \rangle, \vert 1,g \rangle, \vert 2,e \rangle, \ldots  \right\},  
\end{eqnarray}
with $\hat{B}^{\dagger} = (1/\sqrt{\hat{n} +1}) \hat{a}^{\dagger} \hat{\sigma}_{x}$ \cite{Tur2000p574,Tur2001p899,Casanova2010p263603}. These parity bases split the Hilbert space in two orthogonal parity subspaces which are the foundation for the classical simulation of our nonlinear Rabi model.

In addition, if the counter-rotating terms coupling parameter is null, $g_{+}=0$, then the model reduces to a nonlinear JC model that also conserves the total number of excitations defined as $\hat{N} = \hat{n} + \hat{\sigma}_{z} /2$; i.e. $\left[\hat{H}, \hat{N} \right]=0$.
It is simple to obtain the time evolution for this case by using a method that makes use of Susskind-Glogower operators \cite{RodriguezLara2012arXiv1207}, $\hat{V} = \sum_{k=0}^{\infty} \vert k \rangle \langle k+1 \vert$, such that (\ref{eq:OurH}) with $g_{+}=0$ becomes
\begin{eqnarray}
\hat{H} &=& \left( \begin{array}{cc} \hat{V} & 0 \\ 0 & 1 \end{array} \right)  \left( \begin{array}{cc} h(\hat{n} +1) + \omega_{0}/2 & g_{-} f(\hat{n}) \\ g_{-} f(\hat{n}) & h(\hat{n} +1) - \omega_{0}/2 \end{array} \right)  \left( \begin{array}{cc} \hat{V}^{\dagger} & 0 \\ 0 & 1 \end{array} \right) , \\
&=& \left( \begin{array}{cc} \hat{V} & 0 \\ 0 & 1 \end{array} \right) \left( \begin{array}{cc}  \Gamma(\hat{n}) + \Omega(\hat{n}) & \Gamma(\hat{n}) - \Omega(\hat{n}) \\ 2 g_{-} f(\hat{n}) & 2 g_{-} f(\hat{n}) \end{array} \right) \left( \begin{array}{cc} \lambda_{+}(\hat{n}) & 0 \\ 0 & \lambda_{-}(\hat{n}) \end{array} \right) \nonumber \\ && \times  \left( \begin{array}{cc}  \Gamma(\hat{n}) + \Omega(\hat{n}) & \Gamma(\hat{n}) - \Omega(\hat{n}) \\ 2 g_{-} f(\hat{n}) & 2 g_{-} f(\hat{n}) \end{array} \right)^{-1} \left( \begin{array}{cc} \hat{V}^{\dagger} & 0 \\ 0 & 1 \end{array} \right),
\end{eqnarray} 
with the elements of the similarity transformation given by
\begin{eqnarray}
\Gamma(\hat{n}) &=& h(\hat{n}-1) - h(\hat{n}) + \omega_{0}, \\
\Omega(\hat{n}) &=& \sqrt{ \Gamma^{2}(\hat{n}) + 4 g_{-}^{2} f^{2}(\hat{n})},
\end{eqnarray}
and the dispersion relation,
\begin{eqnarray}
\lambda_{\pm} (\hat{n}) = \frac{ h(\hat{n}-1) + h(\hat{n}) \pm \Omega (\hat{n})}{2}.
\end{eqnarray}
Photon transport is then given by the time evolution operator,
\begin{eqnarray}
\hat{U}(t) = &=& \left( \begin{array}{cc} \hat{V} & 0 \\ 0 & 1 \end{array} \right) \left( \begin{array}{cc}  \Gamma(\hat{n}) + \Omega(\hat{n}) & \Gamma(\hat{n}) - \Omega(\hat{n}) \\ 2 g_{-} f(\hat{n}) & 2 g_{-} f(\hat{n}) \end{array} \right) \left( \begin{array}{cc} e^{-i \lambda_{+}(\hat{n}) t} & 0 \\ 0 & e^{- i \lambda_{-}(\hat{n}) t } \end{array} \right) \nonumber \\ && \times  \left( \begin{array}{cc}  \Gamma(\hat{n}) + \Omega(\hat{n}) & \Gamma(\hat{n}) - \Omega(\hat{n}) \\ 2 g_{-} f(\hat{n}) & 2 g_{-} f(\hat{n}) \end{array} \right)^{-1} \left( \begin{array}{cc} \hat{V}^{\dagger} & 0 \\ 0 & 1 \end{array} \right).  
\end{eqnarray}
In other words, by using the analogy between transport of single-photon states and propagation of classical field, it is very simple to calculate the propagation through the equivalent nonlinear JC photonic lattice via quantum optics methods.

\section{Classical simulation in arrays of coupled waveguides}

The dynamics of a Hamiltonian system is given by the Schr\"odinger equation
\begin{eqnarray}
i \partial_{t} \vert \psi(t) \rangle = \hat{H}  \vert \psi(t) \rangle,
\end{eqnarray}
where the notation $\partial_{x}$ stands for partial derivative with respect to $x$.
As our model Hamiltonian (\ref{eq:OurH}) conserves parity, we can split the evolution in even and odd parts, $\vert \psi \rangle = \vert \psi^{(+)}(t) \rangle + \vert \psi^{(-)}(t) \rangle$, and propose the solutions
\begin{eqnarray}
\vert \psi^{(\pm)}(t) \rangle = \sum_{j=0}^{\infty} \mathcal{E}^{(\pm)}_{j}(t) \vert \pm , j \rangle.
\end{eqnarray}
This parity decomposition leads to one coupled differential set for each of the parity subspaces:
\begin{eqnarray}
i \partial_{t} \mathcal{E}^{(\pm)}_{0} &=& d^{(\pm)}(0) \mathcal{E}_{0}^{(\pm)} + g_{\pm} f(1) \mathcal{E}_{1}^{(\pm)}, \label{eq:DSstart} \\
i \partial_{t} \mathcal{E}^{(\pm)}_{2k+1} &=& d^{(\pm)}(2k+1) \mathcal{E}_{2k+1}^{(\pm)} + g_{\pm} f(2k+1) \mathcal{E}_{2k}^{(\pm)} +  g_{\mp} f(2k+2) \mathcal{E}_{2k+2}^{(\pm)}, \quad k \ge 0, \nonumber \\ ~\\
i \partial_{t} \mathcal{E}^{(\pm)}_{2k} &=& d^{(\pm)}(2k) \mathcal{E}_{2k}^{(\pm)} + g_{\mp} f(2k) \mathcal{E}_{2k-1}^{(\pm)} +  g_{\pm} f(2k+1) \mathcal{E}_{2k+1}^{(\pm)}, \qquad \qquad k \ge 1, \nonumber \\
\end{eqnarray}
with 
\begin{eqnarray}
d^{(\pm)}(j) = h(j) \mp (-1)^{j} \frac{\omega_{0}}{2}. \label{eq:DSend}
\end{eqnarray}
Each of these coupled differential sets is equivalent to that describing an array of waveguides coupled to their nearest neighbor up to a global phase and replacing $t \rightarrow z$. 
In this analogy, the refraction index of the $j$th waveguide is modulated by $d^{(\pm)}(j)$ and the separation distance between the $j$th and the preceding waveguide is proportional to $f(j)$.
We are interested in these one-dimensional photonic crystals due to the great control that can be attained in modulating both the refractive index of individual waveguides and the separation between them by femtosecond laser waveguide writing on fused silica \cite{Chan2001p1726}.
In recent experiments, the precise control over the refractive index and coupling parameters of photonic lattices has allowed the classical simulation of the Rabi model for coupling parameters of the order of the field frequency in small photonic lattices \cite{Crespi2012p163601} and, up to our knowledge, it is possible to inscribe as many as sixty waveguides in fused silica with parameters related to Glauber-Fock photonic lattices \cite{Keil2011p103601}, which are of the order required by our nonlinear model, with a present technical limit set in the two hundred inscribed waveguides.
On this account, caution must be exerted in finding the adequate balance between lattice size, parameters of the nonlinear Rabi model, and initial state to be propagated.

One of us has discussed somewhere else \cite{RodriguezLara2011p053845} how a tight binding model of photonic lattices can be solved via matrix methods, $i \partial_{t}  \vec{\mathcal{E}}^{(\pm)} = H^{(\pm)} \vec{\mathcal{E}}^{(\pm)}$ leading to $\vec{\mathcal{E}}^{(\pm)}(t) = e^{- i H^{(\pm)}  t }  \vec{\mathcal{E}}^{(\pm)}(0)$, as long as the functions $d^{(\pm)}(j)$ and  $f(j)$ are time independent. 
In short, the dispersion relation of the truncated parity optical lattices of size $N$ are given by the roots of the characteristic polynomial $p_{N}^{(\pm)}(\lambda)=0$ described by the three-term recurrence relations:
\begin{eqnarray}
p_{0}^{(\pm)}(\lambda) &=& 1, \\
p_{1}^{(\pm)}(\lambda) &=& \left[ d^{(\pm)}(0) - \lambda \right] p_{0}^{(\pm)}(\lambda), \\
p_{2k}^{(\pm)}(\lambda) &=& \left[ d^{(\pm)}(2k-1) - \lambda \right] p_{2k-1}^{(\pm)} - g_{\pm}^{2} f^{2k}(2k) p_{2k-2}, \\
p_{2k+1}^{(\pm)}(\lambda) &=& \left[ d^{(\pm)}(2k) - \lambda \right] p_{2k-1}^{(\pm)} - g_{\mp}^{2} f^{2k}(2k) p_{2k-1}, \quad k \ge 1.
\end{eqnarray}
Note that in the nonlinear JC case, $g_{+} = 0$, its truncation depends heavily on the initial state of the field thanks to the fact that the Hamiltonian conserves the number of excitations. 
In the nonlinear Rabi case the truncation depends heavily on both the initial state and the value of the coupling parameter $g_{+}$, it increases rapidly with the value of $g_{+}$.
For experimental realizations this just means that the lattice must be large enough to keep the propagated classical field far from the last segment of waveguides.

Typical measurements of interest in the quantum optics community include, the mean photon number,
\begin{eqnarray}
\langle \hat{n}(t) \rangle &=& \langle \psi^{(+)}(t) \vert \hat{n} \vert \psi^{(+)}(t) \rangle + \langle \psi^{(-)}(t) \vert \hat{n} \vert \psi^{(-)}(t) \rangle, \\
&\approx& \sum_{j=0}^{N-1} j \left[ \vert \mathcal{E}^{(+)}_{j}(t) \vert^{2} +  \vert \mathcal{E}^{(-)}_{j}(t) \vert^{2} \right] , 
\end{eqnarray}
which is nothing else than the sum of the output intensity at each waveguide weighted by the waveguide position; this quantity has been called the average center of mass of the propagating intensity in other context \cite{Longhi2010p235,Thompson2011p214302,RodriguezLara2013p038116}.
Another relevant quantity is the mean atomic excitation energy,
\begin{eqnarray}
\langle \hat{\sigma}_{z}(t) \rangle &=& \langle \psi^{(+)}(t) \vert  \hat{\sigma}_{z} \vert \psi^{(+)}(t) \rangle + \langle \psi^{(-)}(t) \vert  \hat{\sigma}_{z} \vert \psi^{(-)}(t) \rangle, \\
&\approx& \sum_{j=0}^{(N-1)/2} \left[ \vert \mathcal{E}^{(+)}_{2j + 1}(t) \vert^{2} -  \vert \mathcal{E}^{(+)}_{2j}(t) \vert^{2} \right] + \sum_{j=0}^{(N-1)/2} \left[ \vert \mathcal{E}^{(-)}_{2j}(t) \vert^{2} - \vert \mathcal{E}^{(-)}_{2j + 1}(t) \vert^{2}\right]\nonumber \\ , 
\end{eqnarray}
which is the addition of the intensity on the odd (even) waveguides minus the intensity on the even (odd) waveguides in the positive (negative) parity lattice.
The atomic Berry phase is other quantity of interest, 
\begin{eqnarray}
\langle \hat{\sigma}_{x}(t) \rangle &=& \langle \psi(t) \vert  \hat{\sigma}_{x} \vert \psi(t) \rangle, \\
&\approx& \sum_{j=0}^{N-1} \mathrm{Re} \left[  \left( \mathcal{E}^{(+)}_{j}(t) \right)^{\ast} \mathcal{E}^{(-)}_{j}(t) \right] , 
\end{eqnarray}
which clearly requires knowledge of the relative phases of the field at the outputs as it correlates the field amplitudes of positive and negative parity lattices. 
The fidelity of the time evolved state with respect to the initial state is another quantity of interest that requires knowledge of the relative phases of the amplitudes at the output of the parity lattices, 
\begin{eqnarray}
\mathcal{F} &=& \vert \langle \psi(0) \vert \psi(t) \rangle \vert, \\
&\approx& \left\vert \sum_{j=0}^{N-1} \left( \mathcal{E}^{(+)}_{j}(0) \right)^{\ast} \mathcal{E}^{(+)}_{j}(t) + \left( \mathcal{E}^{(-)}_{j}(0) \right)^{\ast} \mathcal{E}^{(-)}_{j}(t) \right\vert.
\end{eqnarray}
Finally, in order to recover information about the mixedness of the reduced atomic system, it is possible to calculate von Neumann entropy, 
\begin{eqnarray}
\hat{S} = -\mathrm{Tr} \hat{\rho}_{a} ~\mathrm{ln} ~\hat{\rho}_{a}
\end{eqnarray}
where the reduced atomic density matrix is given by
\begin{eqnarray}
\hat{\rho}_{a}(t) &=& \mathrm{Tr}_{f} \vert  \psi(t) \rangle \langle \psi(t) \vert  , \\
& \approx & \sum_{j=0}^{(N-1)/2} \left( \begin{array}{cc}
  \vert \mathcal{E}^{(+)}_{2j + 1} \vert^{2} +  \vert \mathcal{E}^{(-)}_{2j} \vert^{2} &  \mathcal{E}_{2j+1}^{(+)}  \left( \mathcal{E}^{(-)}_{2j+1} \right)^{\ast} + \mathcal{E}_{2j}^{(-)}  \left( \mathcal{E}^{(+)}_{2j} \right)^{\ast} \\
\mathcal{E}_{2j}^{(+)}  \left( \mathcal{E}^{(-)}_{2j} \right)^{\ast} + \mathcal{E}_{2j+1}^{(-)}  \left( \mathcal{E}^{(+)}_{2j+1} \right)^{\ast} &   \vert \mathcal{E}^{(+)}_{2j} \vert^{2} +  \vert \mathcal{E}^{(+)}_{2j} \vert^{2}
\end{array}\right), \nonumber \\
\end{eqnarray}
where the time dependence has been obviated for the sake of space, $\mathcal{E}^{\pm}_{j} \equiv \mathcal{E}^{\pm}_{j}(t)$.

Now, the most common initial state configurations  in quantum optics involve number or coherent states. 
We can simulate a number state times an atomic superposition, 
\begin{eqnarray}
\vert \psi_{F}(0) \rangle &=& c_{e} \vert n, e\rangle + c_{g} \vert n, g\rangle, \\
&=& \left\{ \begin{array}{cl}
c_{g} \vert +, n \rangle +  c_{e} \vert -, n \rangle, & n ~\mathrm{even},\\
c_{e} \vert +, n \rangle +  c_{g} \vert -, n \rangle, & n ~\mathrm{odd}.
\end{array}  \right.
\end{eqnarray}
That is, just one input port is used in each parity lattices. 
In the case of $n$ even the input at the positive and negative parity lattices are $\mathcal{E}_{n}^{(+)} = \vert c_{g} \vert$ and $\mathcal{E}_{n}^{(-)} = e^{i \phi} \vert c_{e} \vert$ with the phase $\phi$ given by the phase difference between $c_{e}$ and $c_{g}$ and the equivalent for $n$ odd.
For coherent states, say the simplest case,
\begin{eqnarray}
\vert \psi_{C}(0) \rangle &=& \vert \alpha, g\rangle, \\
&=& e^{-\vert \alpha \vert / 2} \sum_{n=0}^{\infty} \frac{\alpha^{n}}{\sqrt{n!}} \vert n,g \rangle, \\
&=& e^{-\vert \alpha \vert / 2} \sum_{n=0}^{\infty} \left(  \frac{\alpha^{2n}}{\sqrt{2n!}} \vert +, 2n \rangle + \frac{\alpha^{2n+1}}{\sqrt{(2n+1)!}} \vert -, 2n+1 \rangle \right).
\end{eqnarray}
This state is simply obtained by impinging a field on the $0$th waveguide of a Glauber-Fock lattice \cite{Keil2011p103601, RodriguezLara2011p053845} and propagating it for the corresponding distance.
Then, the output at the even waveguides is sent to the even waveguides of the positive parity lattice and that of the odd waveguides to the odd waveguides of the negative parity lattice.
A highly entangled state involving coherent states can easily be constructed by sending the output from the Glauber-Fock lattice to just the positive or negative parity lattice; for example, if we consider just the positive lattice:
\begin{eqnarray}
e^{-\vert \alpha \vert / 2} \sum_{n=0}^{\infty} \frac{\alpha^{n}}{\sqrt{n!}} \vert +,n \rangle &=& \vert \alpha_{+}, g \rangle + \vert \alpha_{-}, e \rangle ,
\end{eqnarray}
where the even (odd) coherent state is defined by $\vert \alpha_{\pm} \rangle = \vert \alpha \rangle \pm \vert - \alpha \rangle$.
The Glauber-Fock lattice can be used to simulate input corresponding to both coherent and displaced number states.

\begin{figure}[ht]
\center\includegraphics[width=\textwidth]{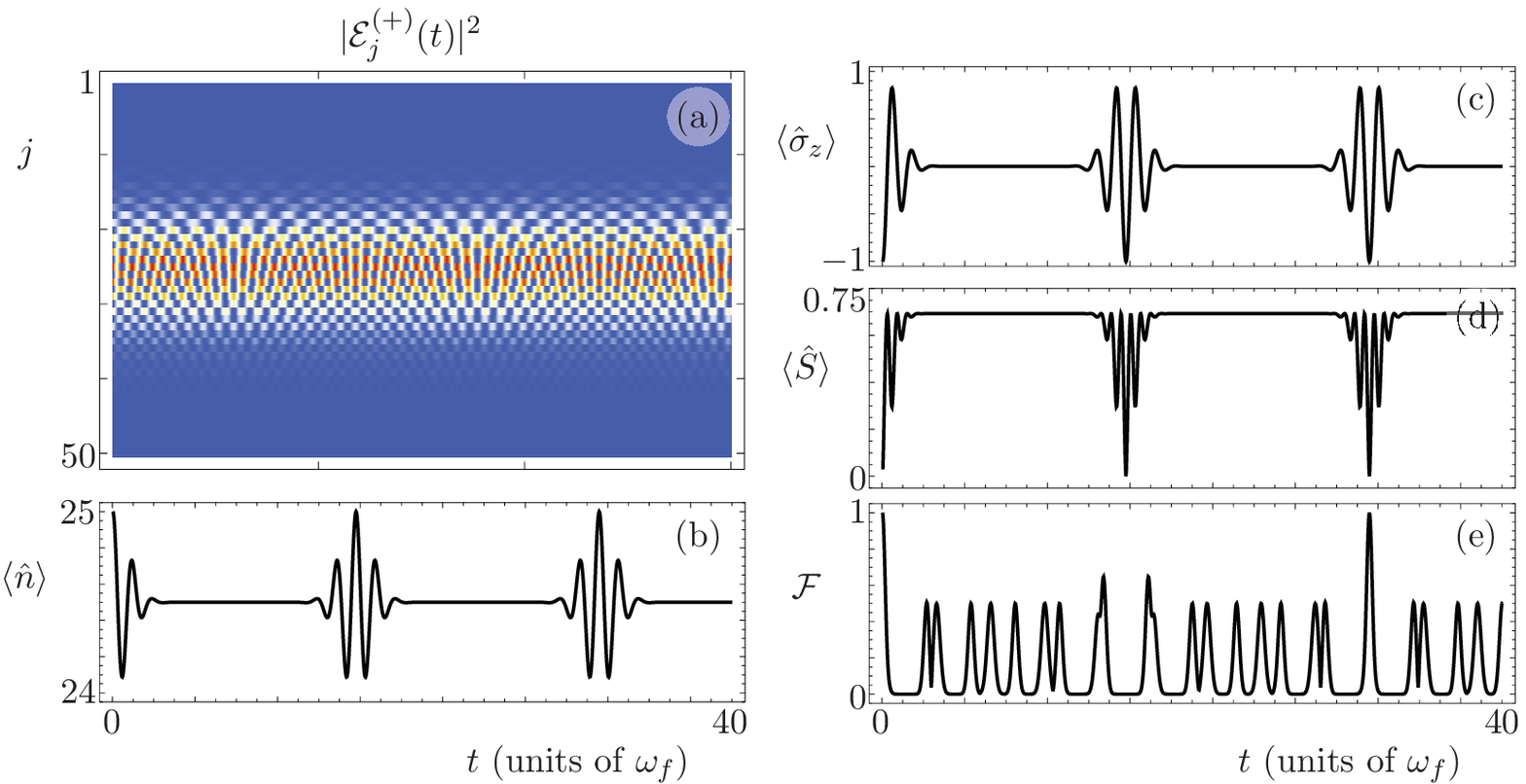}
\caption{ The classical simulation of the time evolution for the separable initial state $\vert \psi(0) \rangle = \vert \alpha_{+}, g \rangle$ with $\alpha = 5$ under BS dynamics on resonance, $\omega_{0}= \omega_{f}$, and coupling parameters $g_{-} = 0.1 \omega_{f}$, $g_{+} = 0$. (a) Propagation of the initial field in the corresponding positive parity photonic lattice of the classical simulator. The time evolution of the (b) mean photon number, (c) mean atomic excitation energy, (d) mean von Neumann entropy, and (e) fidelity reconstructed from the classical simulation. The lattice is composed by three hundred coupled photonic waveguides.}
\label{fig:Fig1}
\end{figure}

\section{An example: The Buck-Sukumar model}

In order to present a practical example, let us refer to the BS model,
\begin{eqnarray}
\hat{H}_{BS} = \omega_{f} \hat{n} + \frac{\omega_{0}}{2} \hat{\sigma_{z}} + g \left( \hat{a} \sqrt{\hat{n}} ~\hat{\sigma}_{+} + \sqrt{\hat{n}} ~\hat{a}^{\dagger} \hat{\sigma}_{-} \right), \label{eq:Hamiltonian}
\end{eqnarray}
that describes the interaction of a two-level system with a field under intensity dependent coupling.  
It is related to our model Hamiltonian (\ref{eq:OurH}) by setting $g_{+} =0$, $g_{-}=g$,  $f(\hat{n})= \hat{n}$ and $h(\hat{n}) = \omega_{f}\hat{n}$.
Thus the positive and negative parity lattices, that classically simulate the BS model, are described by the coupled differential sets in Eq. (\ref{eq:DSstart})-(\ref{eq:DSend}) with the photon number functions $f(\hat{n})= \hat{n}$ and $h(\hat{n}) = \omega_{f}\hat{n}$ and coupling parameters $g_{-} = g$ and $g_{+} = 0$.

\begin{figure}[ht]
\center\includegraphics[width=\textwidth]{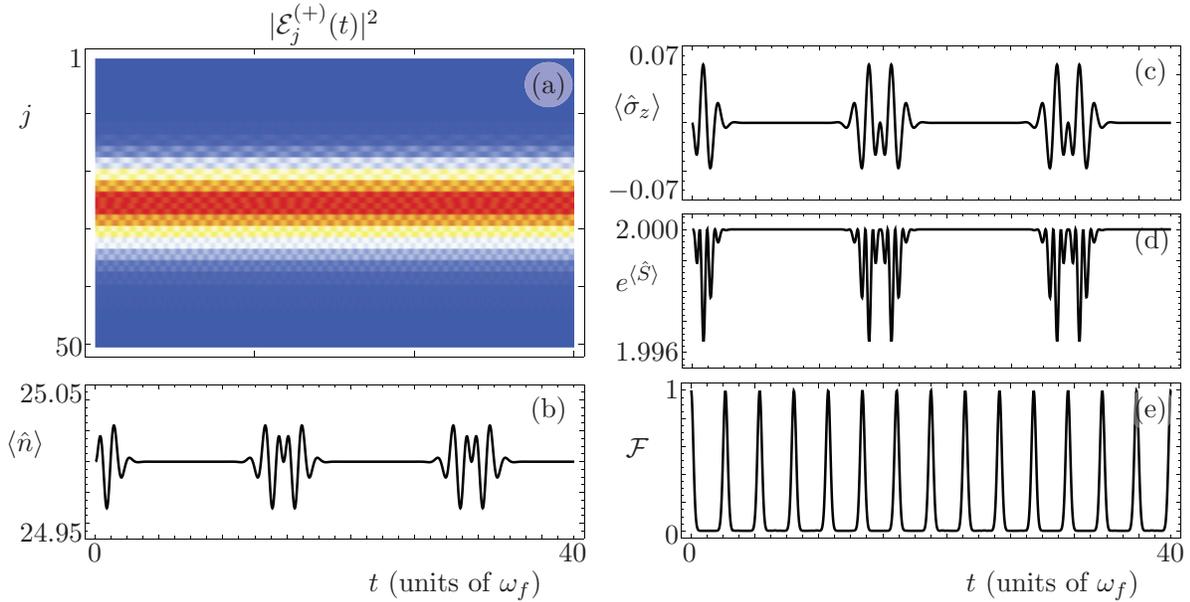}
\caption{ The classical simulation of the time evolution for the entangled initial state $\vert \psi(0) \rangle = \vert \alpha_{+}, g \rangle + \vert \alpha_{-}, e \rangle $ with $\alpha = 5$ under BS dynamics on resonance, $\omega_{0}= \omega_{f}$, and coupling parameters $g_{-} = 0.1 \omega_{f}$, $g_{+} = 0$. (a) Propagation of the initial field in the corresponding positive parity photonic lattice of the classical simulator. The time evolution of the (b) mean photon number, (c)  exponential of the mean atomic excitation energy, (d) mean von Neumann entropy, and (e) fidelity reconstructed from the classical simulation. The lattice is composed by three hundred coupled photonic waveguides. }
\label{fig:Fig2}
\end{figure}

In Figs. \ref{fig:Fig1} and \ref{fig:Fig2}, we show numerical results for the
classical simulation of the BS model on resonance and coupling $g=0.1 \omega_{f}$; i.e. we use a photonic lattice described by the differential sets (\ref{eq:DSstart}-\ref{eq:DSend}) with parameter values $\omega_{0}=\omega_{f}$, $g_{-} = 0.1 \omega_{f}$, and $g_{+}= 0$.
The numerical propagation considers a photonic lattice composed by three hundred coupled waveguides. 
We consider the initial state $\vert \psi(0) \rangle = \vert \alpha_{+}, g \rangle$ with parameter values $\alpha = 5$ in Fig. \ref{fig:Fig1}.
This is a separable state with positive parity; i.e. just the positive parity photonic lattice is needed to classically simulate its evolution.
The intensity of the light field is shown in Fig. \ref{fig:Fig1}(a). 
The time evolution of the mean value for the photon number, atomic excitation energy and von Neumann entropy are shown in Figs. \ref{fig:Fig1}(b)-\ref{fig:Fig1}(d).
Figure \ref{fig:Fig1}(e) shows the time evolution of the fidelity, note how the evolution of this separable initial state returns periodically to its original state.
In Fig. \ref{fig:Fig2}, we show the classical simulation of an entangled initial state 
$\vert \psi(0) \rangle = \vert \alpha_{+}, g \rangle + \vert \alpha_{-}, e \rangle $ with identical parameter values as those in Fig. \ref{fig:Fig1}.
Again, the state has positive parity and just the positive parity photonic lattice is needed to simulate the quantum system.
The evolution of this entangled initial state also returns to its original state as witnessed by the fidelity in Fig. \ref{fig:Fig2}(e).

\begin{figure}[ht]
\center\includegraphics[width=\textwidth]{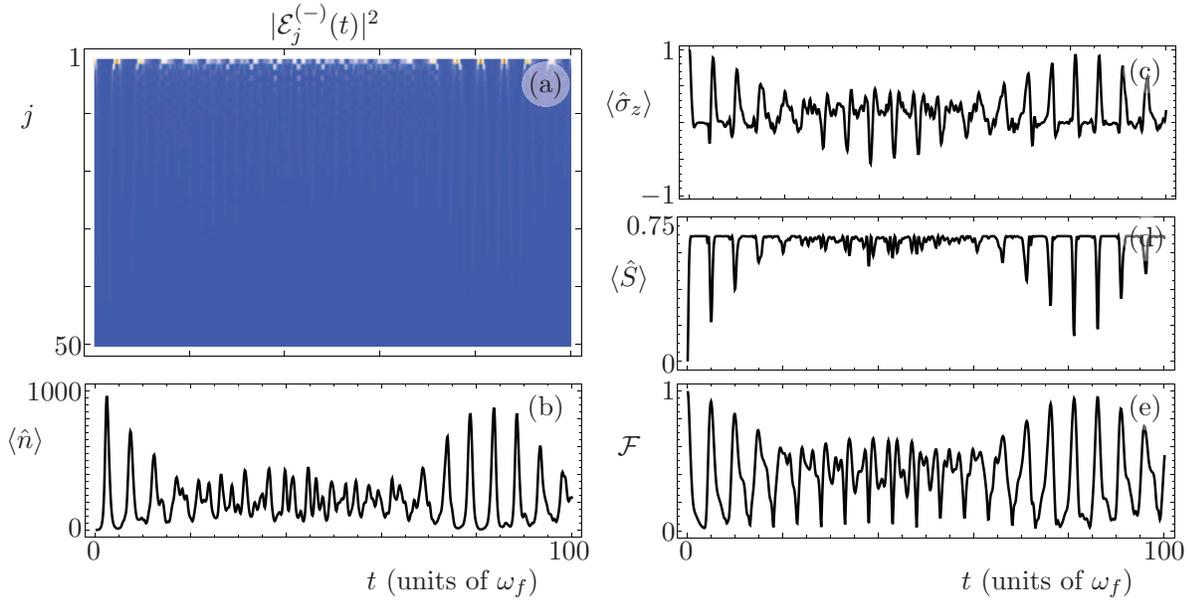}
\caption{ The classical simulation of the time evolution for the separable initial state $\vert \psi(0) \rangle = \vert 0, e \rangle $ under BS plus counter rotating terms dynamics on resonance, $\omega_{0}= \omega_{f}$, and coupling parameters $g_{-} = g_{+} = 2 \omega_{f}$. (a) Propagation of the initial field in the corresponding negative parity photonic lattice of the classical simulator. The time evolution of the (b) mean photon number, (c) mean atomic excitation energy, (d) mean von Neumann entropy, and (e) fidelity reconstructed from the classical simulation. The lattice is composed by two thousand coupled photonic waveguides.}
\label{fig:Fig3}
\end{figure}

For the sake of giving a complete example, we will also consider counter rotating terms,
\begin{eqnarray}
\hat{H} = \omega_{f} \hat{n} + \frac{\omega_{0}}{2} \hat{\sigma_{z}} + g_{-} \left( \hat{a} \sqrt{\hat{n}} ~\hat{\sigma}_{+} + \sqrt{\hat{n}} ~\hat{a}^{\dagger} \hat{\sigma}_{-} \right) + g_{+} \left( \hat{a} \sqrt{\hat{n}} ~\hat{\sigma}_{-} + \sqrt{\hat{n}} ~\hat{a}^{\dagger} \hat{\sigma}_{+} \right). \nonumber \\
\end{eqnarray}
We consider a model on resonance with parameter values $\omega_{0} = \omega_{f}$ and $g_{-} = g_{+} = 2 \omega_{f}$.
Figure \ref{fig:Fig3} shows the numerical results for the propagation of a light field that simulates the initial state $\vert \psi(0) \rangle = \vert 0, e \rangle$ which has negative parity and corresponds to impinging the first waveguide of the negative parity photonic lattice. 
Again, the intensity of the light field is shown in Fig. \ref{fig:Fig3}(a). 
The time evolution of the mean value for the photon number, atomic excitation energy and von Neumann entropy are shown in Figs. \ref{fig:Fig3}(b)-\ref{fig:Fig3}(d).
The numerical propagation considers a photonic lattice of size two thousand and the probability of finding light at the last waveguide has a maximum value of $7\times10^{-4}$ within the parameter range considered here.
Note that this is a thought experiment at the time because current technology can produce a couple hundred coupled waveguides at most. 

\section{Conclusion}

We showed how the parity of the nonlinear Rabi model in (\ref{eq:OurH}), which is unphysical for most parameter sets with current quantum technology, can be exploited to simulate its dynamics by propagating classical light fields in a set of two photonic lattices, each one corresponding to one of the parity subspaces. 
We discussed how initial states in the quantum system involving Fock states map to impinging the photonic lattices at specific waveguides.
Initial states involving coherent or number displaced states can be engineered by propagation through Glauber-Fock lattices.
As an example, we presented numerical results for the classical simulation of the Buck-Sukumar model with initial states within the positive parity subspace describing fully separable and maximally entangled states. 
Also, we considered the Buck-Sukumar model including counter-rotating terms for an initial state in the negative parity subspace.

\section*{Acknowledgment}

HMMC is grateful to Demetrios N. Christodoulides for his hospitality and acknowledges support from the sabbatical leave program of CONACYT.

%

\end{document}